\documentclass{article}

\usepackage{amsmath, amsthm, amssymb, subfigure, wrapfig, verbatim}
\usepackage[T1]{fontenc}
\usepackage{color}
\usepackage{graphicx}

\newtheorem{theorem}{Theorem}
\newtheorem{lemma}{Lemma}
\newtheorem{definition}{Definition}

\newtheorem{corollary}{Corollary}

\newcommand{\john}[1]{\marginpar{\color{blue}$XXX$}\textsc{\color{blue}John proclaims:} \textsl{#1}}
\newcommand{\greg}[1]{\marginpar{\color{red}$XXX$}\textsc{\color{red}Greg proclaims:} \textsl{#1}}
\newcommand{\erik}[1]{\marginpar{\color{Orange}$XXX$}\textsc{\color{Orange}Erik proclaims:} \textsl{#1}}
\newcommand{\vida}[1]{\marginpar{\color{Bittersweet}$XXX$}\textsc{\color{Bittersweet}Vida proclaims:} \textsl{#1}}
\newcommand{\cd}[2]{#1 \cdot #2}

\title{Minimum Feature Size Preserving Decompositions}

\author{%
Greg Aloupis\thanks
{Universit\'e Libre de Bruxelles, Brussels, Belgium. aloupis.greg@gmail.com}
\and
  Erik D. Demaine\thanks
  {  Massachusetts Institute of Technology, Cambridge, Massachusetts, USA. edemaine@mit.edu
}
\and
  Martin L. Demaine\thanks
  {  Massachusetts Institute of Technology, Cambridge, Massachusetts, USA. mdemaine@mit.edu
}
\and 
 Vida Dujmovi\'{c}\thanks
 { Carleton University, Ottawa, Ontario, Canada. vida@cs.mcgill.ca
}
\and
  John Iacono\thanks
  {
    Polytechnic Institute of New York University, Brooklyn, New York, USA. jiacono@poly.edu
}
   }

\newtheorem{construction}[theorem]{Construction}

\newcommand{\irmfs}{internal minimum feature size}
\newcommand{\rmfs}{minimum feature size}
\newcommand{\mfs}{spread}

\DeclareMathOperator{\mmfs}{spread}
\DeclareMathOperator{\mrmfs}{mfs}
\DeclareMathOperator{\mdiam}{diam}

\date{}

\begin{document}
\maketitle

\begin{abstract}
The \emph{minimum feature size} of a crossing-free straight line drawing
is the minimum distance between a vertex and a non-incident edge. This quantity
measures the resolution needed to display a figure or the tool size needed
to mill the figure.
The \emph{spread} is the ratio of the diameter to the minimum feature size.
While many algorithms (particularly in meshing) depend on the spread of
the input,  none explicitly consider finding a mesh whose
spread is similar to the input.
When a polygon is partitioned into smaller regions, such as triangles or quadrangles, the {\em degradation} is the ratio of original to final spread (the final spread is always greater).

\indent Here we present an algorithm to quadrangulate a simple $n$-gon, while achieving
constant degradation.  Note that although all faces have a quadrangular shape, the number of edges bounding each face may be larger. This method uses $\Theta(n)$ Steiner points and produces $\Theta(n)$ quadrangles.   In fact to obtain constant degradation, $\Omega(n)$
Steiner points are required by any algorithm.

We also show that, for some polygons, a constant factor cannot be achieved by any
triangulation, even with an unbounded number of Steiner points. The specific lower bounds depend on whether Steiner vertices are used or not.
\end{abstract}

\section{Introduction}

\subsection{Problem and Motivation}

This paper contains an analysis of  planar polygon decompositions, focusing on 
the objective that among all new point-edge distances produced, none should be much smaller than those in the input.
Specifically, the \emph{minimum feature size} of a noncrossing planar straight
line drawing (a special case of which is the noncrossing set of line segments forming a
polygon) is the minimum distance between two non-touching edges.
Our goal is to bound the change in minimum feature size that results from decomposing a polygon into triangles, quadrangles, and so on.

Feature size was introduced in the context of meshing \cite{Ruppert-1993},
where it influences the necessary mesh complexity that guarantees certain levels of quality.
See also~\cite{Dey-2007,Hudson-Miller-Phillips-2006}.

Another motivating example is the 95-year-old algorithm by
Lowry \cite{Lowry-1814,Frederickson-1997} for finding a common dissection 
of any two polygons of equal area.  This algorithm starts by triangulating
the polygon, then using a dissection from 1778 to convert each triangle
into a rectangle with a common height $\epsilon$ equal to half the minimum
height of all triangles.  The algorithm uses $O(A/\epsilon)$ pieces
where $A$ is the area of the polygon, and this bound is clearly the best
possible for each individual piece.  The issue is that $\epsilon$
depends on the choice of triangulation. We show
that a triangulation without Steiner points could be forced to have $\epsilon$
arbitrarily small, even if the input polygon has a constant number of
vertices and a constant {\em spread}.
The spread is the ratio of {\em diameter} over minimum feature
size---roughly, the ratio between the largest and smallest distances.
We remind the reader that the diameter is the largest distance between any two points in the union of the edges. The term ``spread'' comes from the analogous measure for point sets
~\cite{Erickson-2003}.

%
Essentially, this paper can be thought of as an implicit study of meshing 
with the objective of  preserving feature size,
or equivalently, spread  (meshing does not affect diameter).\\
\\
Lowry's algorithm is an example of a {\em pseudopolynomial bound}.
Quite common in
problems with integer inputs,
pseudopolynomial bounds are polynomial in the input size, $n$,
and in the sum of the input integers, often denoted~$N$.
Subset Sum, for example, has a pseudopolynomial-time algorithm,
but has no polynomial-time algorithm unless P = NP,
being weakly NP-complete \cite{Garey-Johnson-1979}.

Pseudopolynomial bounds are also common in computational geometry,
where problems have real inputs such as point coordinates.
One approach is to assume that the real inputs are rationals,
which includes most representations of real numbers on a digital computer,
scale them all to become integers,
and let $N$ be the sum of the resulting integers.
However, some geometries, especially the outputs of geometric algorithms,
are impossible to represent with rational coordinates.
For this reason, computational geometers generally prefer the
\emph{real RAM} model of computation, where each input can be any real number,
and algorithms have the ability to do only basic arithmetic, radical, and
sign computations on these real numbers.

The usual approach for a pseudopolynomial bound on a real RAM is to 
measure the effective resolution of the input using basic geometric quantities.
A pseudopolynomial bound might be polynomial in the input size $n$
and in the spread~$r$.
Note that, as with problems with (distinct) integers, $n \leq N$,
here we have a (slightly weaker) bound of $n = O(r^2)$.

While many algorithms attain pseudopolynomial bounds, e.g., on running time,
and thus depend on spread, few algorithms consider the spread of their output, as we have seen in our example.
This omission becomes important when trying to chain algorithms together.
Thus one can see the importance of our objective of obtaining a bounded increase in feature size when producing a decomposition.

\subsection{Preliminaries}

\begin{definition}
\label{def-mfs}
 For any planar straight line drawing $G$, the \emph{\rmfs},  
$\mrmfs(G)$, is the minimum distance between a vertex and a non-incident edge. 
Let $\mdiam(G)$ denote the diameter of $G$. The \mfs, $\mmfs(G)$, is $\frac{\mrmfs(G)}{\mdiam(G)}$.
\end{definition}

We are interested in subdividing an $n$-sided polygon $P$ into a planar straight-line drawing $G$ such that the \rmfs\ of $G$ is as close as possible to that of  $P$. We call the ratio $\frac{\mrmfs(P)}{\mrmfs(G)}$ the \emph{degradation} of the decomposition of $P$ into $G$.
 In this paper we bound the degradation in terms of $n$, for various common types of decompositions.

\begin{figure}[htb]
\center\includegraphics[width=5in]{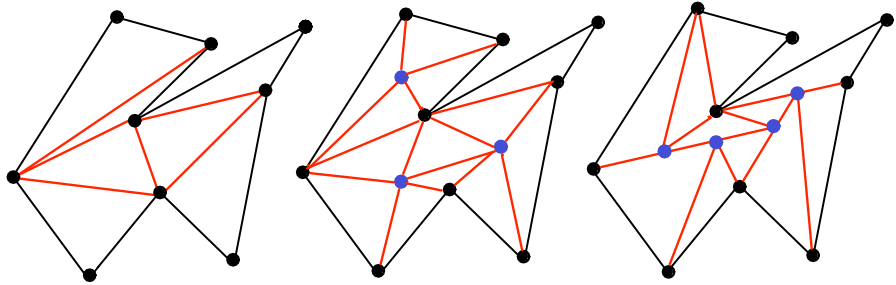}
\caption{Types of triangulations: classic, proper, non-proper. Steiner points are blue.}
\label{triangulationtypes-fig}
\end{figure}

We will be looking at three types of decompositions, known as \emph{triangulations}, where the interior of $P$ is partitioned into triangles. 
See Figure~\ref{triangulationtypes-fig}.
The most common decomposition of a polygon is the \emph{classic triangulation}, where non-crossing chords are added between vertices of $P$, until the interior of $P$ is partitioned into triangles.  More general types of triangulations allow the addition of vertices (known as {\em Steiner} points) on or inside $P$.  We will consider two such triangulations;  a \emph{proper} triangulation forbids the
placement of Steiner points in the interior of any edge. In other words, no two edges incident to a common Steiner point are collinear.  In the dual graph of the triangulation all vertices have degree at most 3.  A {\em non-proper} triangulation simply partitions $P$ into triangles, with no restrictions.  Three types of \emph{quadrangulations}, or in general $k$-rangulations, are defined analogously.

\subsection{Contributions in this paper}

In section~\ref{noclassic} we show that classic $k$-rangulations, for any $k$, cannot guarantee low degradation of \rmfs. Specifically, we show all classic $k$-rangulations of a regular $n$-gon have a \rmfs\ degradation of $\Omega(n)$.
In Section~\ref{properlb} we show that there is a family of $n$-sided polygons for which  all proper $k$-rangulations  have a \rmfs\ degradation of  $\Omega \left( \frac{\log n}{\log \log n} \right)$. 
On the positive side,  in section~\ref{bigalg} we give a
 $\Theta(n)$-time algorithm that produces a $\Theta(1)$-degradation non-proper quadrangulation
for any polygon with $n$ sides.
 %
 

\begin{table}
\begin{center}
\begin{tabular}{c|c|c}
 & Triangulation & Quadrangulation \\ \hline
Classic & $\Omega(n)$ & $\Omega(n)$ \\
Proper & $\Omega \left( \frac{\log n}{\log \log n} \right), O(\log n)$ & $\Omega \left( \frac{\log n}{\log \log n} \right),O(\log n)$\\
Non-proper & $O(\log n)$ & $\Theta(1)$\\
\end{tabular}
\vspace{1pc}
\caption{\textbf{Our Results.} The best \rmfs\ degradation values are indicated, for various
decompositions of an $n$-vertex  polygon. The quadrangulation results also hold for any $k$-rangulation for $k\geq 4$. We conjecture that the three $O(\log n)$ values  are  tight.
}
\end{center}
\end{table}


\section{Lower bound for classic triangulations} \label{noclassic}
\label{sec-steiner-needed}

\begin{lemma} \label{regpoly}
For every $n$, there is a polygon $P_n$ such that  all classic triangulations of $P_n$ have degradation $\Omega(n)$.
\end{lemma}

\begin{proof}


Consider a regular even $n$-gon $P_n$, with unit sides. The \rmfs\ is 1.   Thus the \mfs\ is $\sin \frac{\pi}{n}$.

By the {\em Two Ears Theorem}, any classic  triangulation $T$ of a polygon must contain two disjoint triangles, each containing two adjacent  edges of  the polygon.
The \rmfs\ of these triangles is $\sin \frac{\pi(n-2)}{2n}$.  Thus any  triangulation of
$P_n$ will have a degradation of $\Omega(n)$.
%
 \end{proof}

The arguments of the above proof  can be extended trivially to quadrangles or any decomposition
of constant size.

\section{Bounds on proper triangulations}
\label{properlb}
In this section we prove that any constant-size decomposition (i.e. $r$-rangulation)
of a certain class of $n$-gons must have a degradation of $\Omega(\frac{\log n}{\log\log n})$.
We first bound the maximum degree that a vertex can have within a rectangle, in order
to maintain a unit \rmfs\ (section~\ref{maxdegrec}).  The main result follows in section~\ref{sec:logloglogproof}.

\subsection{Maximum Degree in a Rectangle}
\label{maxdegrec}

\begin{figure}[h!tb]
\begin{center}
\includegraphics[width=1.7in]{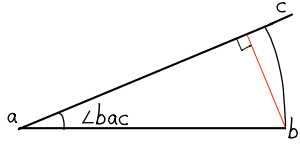}
\end{center}
\caption{Proof of Lemma~\ref{l:lbangle}.}
\label{lbangle}
\end{figure}

\begin{lemma}
\label{l:lbangle}
Given vertex $a$ and two incident edges $\overline{ab}$ and $\overline{ac}$, if \rmfs\ $=1$, then
$\cd{\angle bac} {\min (|ab|,|ac|)} \geq 1$.
\end{lemma}
\begin{proof}
Refer to Figure~\ref{lbangle}.
Assume w.l.o.g.~that $1\leq |ab| \leq |ac|$. The distance from $b$ to $\overline{ac}$ is  $ |ab| \cdot \sin \angle{bac}$. This distance must be at least 1 in order to achieve unit \rmfs. 
Since $|ab| \cdot \angle bac \geq |ab|\cdot \sin \angle bac$, our claim holds.

\end{proof}

\begin{definition}
For a point  $c$ in a
rectangle $R$, let $d(c,\theta,R)$ be the distance from $c$ to the boundary of $R$ along a ray of angle $\theta$ with respect to the $x$-axis.
\end{definition}

\noindent The following lemma relies on elementary trigonometry and thus no proof is given.
\begin{lemma}
\label{proofless}
Let  $c,p$ be two points in a $k \times n$ axis-aligned rectangle $R$.
Let $\theta$ be the angle between $\overline{cp}$ and the $x$-axis.
If $\theta \leq \tan^{-1}\frac{n}{k}$ then
$d(c,\theta,R) \leq \frac{k}{\cos \theta}$. If $\theta \geq \tan^{-1}\frac{n}{k}$ then $d(c,\theta,R) \leq \frac{n}{\sin \theta}$.
\end{lemma}

\begin{theorem}\label{degree}
Within  a $k \times n$ rectangle $R$, let a point $c$ 
be connected to $q$ points $\{p_1,\ldots,p_q\}$.
The \rmfs\ can equal 1 only if  $q = O(k \log \frac{n}{k})$.
\end{theorem}

\begin{proof}
Assume w.l.o.g.~that  $R$ is axis-aligned and at least $q/4$ of the points $p_i$ are 
above and to the right of $c$.  Let this subset be denoted by $P'$.  
Let $\gamma_i$ be the angle of $\overline{cp_i}$ relative to the $x$ axis.
Separate the points in $P'$ into two groups $P^{\ell},P^s$ depending on whether $\gamma_i$ is 
larger or smaller (respectively) than $\tan^{-1} \frac{n}{k}$.  Let $P''$ be the larger of the two groups.  Let $l = |P''|$ and assume that the points in $P$ are numbered such that $P''{=}\{p_1, p_2, \ldots p_l\}$. Furthermore for $1 \leq i <j \leq l$, if $P''=P^s$, $\gamma_i \leq \gamma_j$ 
(otherwise  if $P''=P^{\ell}$, $\gamma_i \geq \gamma_j$).

We claim that $\int_{\gamma_i}^{\gamma_{i+1}} {d(o,\theta,R)}d \theta \geq 1$, where $o$ is the  lower-left corner of $R$.

\noindent By Lemma~\ref{l:lbangle}:

$$  1 \leq \cd{\angle p_i c p_{i+1}}{\min(|cp_i|,|cp_{i+1}|)} $$

\noindent Since for all $i$ the segment $cp_i$ is inside $R$, $|cp_i| \leq d(c,\gamma_i,R)$, so

$$ \leq \cd{\angle p_i c p_{i+1}}{\min(d(c,\gamma_i,R),d(c,\gamma_{i+1},R))} $$

\noindent For any given angle within the range considered here, the distance from $o$ to the boundary of $R$ is greater than the equivalent distance from $c$. Thus we have:

$$ \leq \cd{\angle p_i c p_{i+1}}{\min(d(o,\gamma_i,R),d(o,\gamma_{i+1},R))} $$

\noindent Since the distances increase with the angles:

$$ = \cd{\angle p_i c p_{i+1}}{d(o,\gamma_{i+1},R)} $$

\noindent Since $\angle p_i c p_{i+1}= \gamma_{i+1}-\gamma_i$:

$$ = \int_{\gamma_i}^{\gamma_{i+1}} {d(o,\gamma_{i+1},R)} d \theta $$

\noindent Again, because distances increase with angles:

$$ \leq \int_{\gamma_i}^{\gamma_{i+1}} {d(o,\theta,R)}d \theta $$

\noindent
Thus we establish our claim that  $1 \leq  \int_{\gamma_i}^{\gamma_{i+1}} {d(o,\theta,R)}d \theta $. Next we claim that the following holds:

$$ 8(q{-}1) \leq l =  \sum_{i=1}^l 1\leq  \sum_{i=1}^{l-1} \int_{\gamma_i}^{\gamma_{i+1}}{d(c,\theta,R)}d \theta $$

$$ \leq   \int_{\gamma_1}^{\gamma_{i}} {d(o,\theta,R)}d \theta $$

\noindent The proof depends on whether  $P'' = P^{\ell}$ (left) or $P''=P^s$ (right).
In the former case, angles  range from $0$ to $\tan^{-1}\frac{n}{k}$,  whereas in the latter case they range from  $\tan^{-1}\frac{n}{k}$ to $\frac{\pi}{2}$.

$$ \leq   \int_{0}^{\tan^{-1}\frac{n}{k}} {d(o,\theta,R)}d \theta \hspace{3pc}
\leq   \int^{\frac{\pi}{2}}_{\tan^{-1}\frac{n}{k}} {d(o,\theta,R)}d \theta $$

\noindent 
By Lemma~\ref{proofless}, the left and right inequalities respectively become:

$$ =   \int_{0}^{\tan^{-1}\frac{n}{k}} \frac{k}{\cos \theta}d \theta \hspace{3pc} =
 \int^{\frac{\pi}{2}}_{\tan^{-1}\frac{n}{k}} \frac{n}{\sin \theta}d \theta $$

\noindent The right expression is simpler to bound so we handle it first.
Since  $n{+}k$ is a trivial upper bound on any distance in $R$,

 $$\leq \int^{\frac{\pi}{2}}_{\tan^{-1}\frac{n}{k}} (n{+}k)d \theta $$

\noindent Having removed the dependency on  $\theta$, we manipulate the limits of the integral and use $\tan^{-1} x = \frac{\pi}{2}-\tan^{-1} x^{-1}$ to obtain:

 $$= \int_{0}^{\tan^{-1} \frac{k}{n}} (n{+}k)d \theta = (n{+}k)\tan^{-1} \frac{k}{n} $$

\noindent Since $\tan^{-1} x= x - \frac{x^3}{3}+ \frac{x^5}{5}- \frac{x^7}{7}- \frac{x^9}{9} \ldots $, and $n\geq k$, we get

$$ = O\left( (n{+}k) \frac{k}{n} \right) = O(k)$$

\noindent This concludes the right case. We return to the left case, 
$\int_0^{\tan^{-1} \frac{n}{k}} \frac{k}{\cos \theta} d\theta $.
Since $\int\frac{1}{\cos \theta}d\theta = 2 \tanh^{-1} \tan \frac{\theta}{2}$:

$$ = 2 k \tanh^{-1} \left( \tan \frac{\tan^{-1} \frac{n}{k}}{2} \right) - 2 k \tanh^{-1} \left( \tan \frac{0}{2} \right)$$

\noindent Since $\tanh^{-1} \tan 0=0$:

$$=  2 k  \tanh^{-1}  \tan \left( \frac{1}{2}\tan^{-1} \frac{n}{k} \right)$$

\noindent Since $\tanh^{-1} x = \frac{1}{2}(\log (1{+}x) - \log (1{-}x))$:

$$=  2 k \frac{1}{2} \left( \log  \left(1+ \tan \left( \frac{1}{2}\tan^{-1} \frac{n}{k} \right) \right) 
- \log \left( 1- \tan \left( \frac{1}{2}\tan^{-1} \frac{n}{k} \right) \right) \right)$$

\noindent Since $\tan \frac{1}{2}x \leq \tan x$ for $x \geq 0$:

$$ \leq  k  \left( \log  \left(1+ \tan \tan^{-1} \frac{n}{k}  \right) 
- \log \left( 1-\tan \left( \frac{1}{2}\tan^{-1} \frac{n}{k} \right) \right) \right)$$


$$ = k  \left( \log  \left( 1+\frac{n}{k}  \right) 
- \log \left( 1-\tan \left(  \frac{1}{2}\tan^{-1} \frac{n}{k} \right) \right) \right)$$

The value $ \tan (\frac{1}{2}\tan^{-1} \frac{n}{k})$ ranges between 0 and 1.  However it equals 1
only if $k=0$, which is a degenerate case that we exclude.  Therefore the second $\log$ term is
well defined and always has a negative value. 
Since  $n\geq k$, the negative value is bounded by a constant, so we can claim
that:



$$ \leq k  \left( \log  \left( 1+\frac{n}{k}  \right) 
+ O(1) \right)  = O \left( k \log \frac{n}{k} \right) $$

Combining the $O(k)$ and $O \left( k \log \frac{n}{k} \right)$ bounds for the left and right cases yields  $$ q = O \left( k \log \frac{n}{k} \right) $$

\end{proof}

\subsection{Main Proof}
\label{sec:logloglogproof}

\begin{lemma}
\label{parentlemma}
Let $T$ be a proper $r$-rangulation of polygon $P$. Let $p$ be a vertex such that the leftward ray from $p$ starts out within $P$ and intersects some edge of $T$.
Then $p$ has a path of at most $r{-}2$ edges to a vertex at least $1$ unit to its left. 
\end{lemma}

\begin{figure}[h!]
\begin{center}
\includegraphics[width=1.7in]{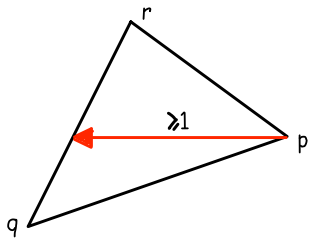}
\end{center}
\caption{Illustration of Lemma~\ref{parentlemma}.}
\label{pickparent}
\end{figure}

\begin{proof}
Refer  to Figure~\ref{pickparent}.  Let $rq$ be the first segment crossed by the leftward ray from $p$ in $T$. Assume w.l.o.g.~that $q$ is to the left of $r$. Note that since the \rmfs\ of $T$ is at least 1,  the ray must travel at least a unit distance before hitting $rq$.  Thus the horizontal distance from $p$ to $q$ is at least 1. 
 
If $T$ is a proper triangulation then $p$ must be connected to $q$ by an edge in $T$, as illustrated on the right in Figure~\ref{pickparent}. In general, if $T$ is a proper $r$-rangulation, $q$ and $p$ must still be on the same face and are connected by a chain of at most $r{-}2$ edges.
Note that this is not true in a non-proper $k$-rangulation.

\end{proof}

Clearly, Lemma~\ref{parentlemma} holds regardless of the direction chosen for the ray.
In general for a given direction, we say that the selected vertex $q$ is the {\em parent} of $p$.

\begin{figure}[h!]
\begin{center}
\includegraphics[width=4in]{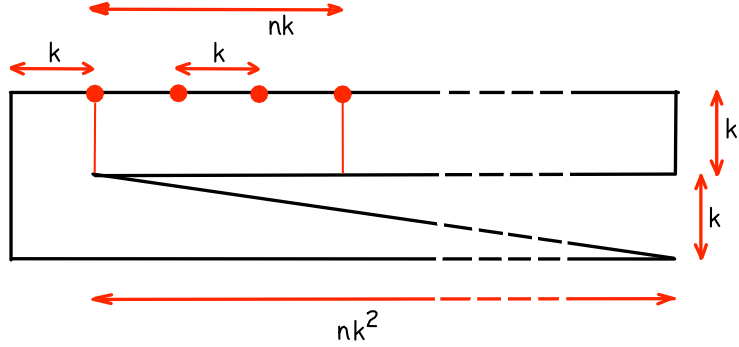}
\end{center}
\caption{Lower bound example for proper $r$-rangulations: polygon $P(n,k)$.}
\label{lbproper}
\end{figure}

Given positive integers $n$ and $k$, let $P(n,k)$ be the polygon illustrated in Figure~\ref{lbproper}.  If its lower-left vertex is the origin, the five remaining vertices that define its shape have 
coordinates $(0,2k)$, $(k{+}nk^2, 2k)$,  $(k{+}nk^2, k)$,  $(k, k)$,  $(k{+}nk^2, 0)$.
Thus the polygon is contained in a $(k{+}nk^2) \times 2k$ rectangle.
An additional $n$ vertices are placed $k$ units apart from each other, starting from the top-left
vertex on the top horizontal edge of $P(n,k)$.

The \rmfs\ of $P(n,k)$ is $k$.  We define the rectangle $R=[(k,k), (k,2k), (k{+}nk,2k), (k{+}nk, k)]$ as the
{\em critical region}.

\begin{theorem}
\label{logloglog}
If $P(n,k)$ has a proper $r$-rangular decomposition  with \rmfs\ 1,
then $k = \Omega \left( \frac{\log n}{\log \log n}\right)$.
\end{theorem}

\begin{proof}
First, observe that no Steiner points can be placed in the interior of the bottom edge of the critical region $R$, because they would be less than one unit away from the diagonal edge.

Since the decomposition is proper, each of the $n$ vertices at the top of $R$ must be
incident to chords.  These chords may end at the lower two vertices of $R$, or at Steiner points
within $R$, or may lead outside of $R$ via its left/right edges.

Notice that at most $k{-}1$ chords can cross either of the left/right edges of $R$, since the
length of those edges  is $k$. 

Now consider  the set of vertices in $R$, including any Steiner vertices added in a $r$-rangular
decomposition.  For each of these vertices, we choose the downward direction to construct
parent relations, according to Lemma~\ref{parentlemma}.  The only vertices that are not assigned
parents are the two located on the bottom edge of $R$.  Thus the ``sinks", i.e. vertices where the parent hierarchy ends, are the two bottom vertices and/or vertices outside $R$.   These relations form a forest
of trees, with a total of $n$ leaves on the top edge of $R$.  Note that the number of trees is at
most $2k$ since we have established that this is the number of chords that can exit $R$.
Thus one of these trees , $T$, must have at least $\frac{n}{2k}$ leaves.

According to Theorem~\ref{degree}, in a  decomposition of a $nk \times k$ rectangle, the degree of any vertex  must be $O\left( k\log \frac{nk}{k} \right) = O(k\log n)$ in order to preserve a unit \rmfs.
This implies that $T$ has height $\Omega(\log_{k\log n}(n/2k))$.  Furthermore,
the same bound holds for the vertical distance from root to leaf, since each parent pointer 
represents a vertical separation of at least one unit, spread over at most $r{-}2$ vertices 
(where $r$ is a constant).    

However, by construction we know that this vertical distance is at most $k$.  
So we obtain $k = O(\log_{k\log n}(n/2k))$.   
%
%

\end{proof}

Theorem~\ref{logloglog} tells us that by selecting $k= \frac{\log n}{\log \log n}$, 
the \rmfs\ will drop from $k$ to 1.  Thus we obtain the following.
\begin{corollary}
For any constant $r$, there exist $n$-gons for which all proper $r$-rangulations have a degradation of 
$\Omega \left( \frac{\log n}{\log \log n}\right)$.
\end{corollary}

\section{Non-proper quadrangulations can preserve \rmfs}
\label{bigalg}

In this section we show how to construct a non-proper quadrangulation for any polygon,
such that the \rmfs\ degradation is $\Theta(1)$.  We use $\Theta(n)$ Steiner points, and the construction can be computed in linear time.\\

Let  $P{=}\{p_1,\ldots,p_n\}$ be an $n$-gon with  \rmfs\ 1.  
Consider the well-known {\em grassfire transformation}, $G_P(t)$, commonly used to visualize the
formation of the medial axis of a polygon $P$. $G_P(t)$ is the result of shrinking $P$ by
lighting a fire along its boundary, and assuming that the fire will progress at unit speed and will
last for $t$ units of time.
If $P$ has unit  \rmfs, $G_P(t)$ will produce a connected region as long as $t<0.5$.

We construct the curve $P_2 =
 G_P(\frac{2}{5})$. 
Once more we apply the grassfire transformation, this time to $P_2$, but we
expand outward.   In some sense we are computing $P_1 = G_{P_2}^{-1}(\frac{1}{5})$, where the inverse sign denotes outward expansion.
It is  not difficult to see that $P_1$ is contained in $G_P(\frac{1}{5})$.   
Thus we have constructed a ``tube" of constant\footnote{Here, constant width means that every point on the outer boundary has distance 1/5 to some point on the inner boundary, and vice versa. In fact, only reflex vertices, where the tube bends, have multiple equidistant points (along circular arcs).} 
width 1/5, confined between 
$G_P(\frac{1}{5})$ and $G_P(\frac{2}{5})$.

Our tube has the property that any point inside it has distance at least 1/5 to $P$.
This is important because we intend to place Steiner vertices in the tube, specifically in the 
vicinity of each bend.  In fact we will construct a polygon with boundary contained in the tube.

The tube consists of $n$ bends, one per vertex of $P$.  Each bend consists of 
a ``greater" circular arc (spanning an angle of at most $\pi$), a ``minor" co-circular arc (possibly
a degenerate point, spanning the same angle at constant distance 1/5 from the greater arc), and two segments of length 1/5, joining the arc endpoints (at right angles).   
Let us call the two segments {\em doors}.
Between each bend (from door to door), the tube has a rectangular shape of width 1/5.

We will loosely construct our (Steiner) polygon, by providing a circular region for the location of each vertex.
These circles will have a diameter of 1/20, i.e. a quarter of the tube's width.
Later on we will fix the precise location of each vertex in its region.
%
Each circle will have its center on a specific curve which we call the tube's {\em track}.
The track is constructed as follows. Within each bend, from door to door, the track is co-circular to the bend at a constant distance of 1/40 from the greater arc.
Between two bends, the track is a straight connecting segment.

Now we can fill the track with circles, evenly spacing them so that two consecutive circles are
1/20 apart.  The first and last circles added may be 3/20 apart, but then we can re-position everything and the spacing will become slightly larger, without affecting our claims.

A ``greater" arc has length at most $\frac{2}{5}\pi$, so the track within one bend has length
at most $\frac{15}{40}\pi$.   Thus the number of circles intersecting each bend is constant
(roughly 25, at most. In fact this can probably be reduced to around 3-4).
We delete all circles not intersecting a bend.
 Thus the total number of circles (and Steiner vertices)
will be $O(n)$.

Now we superimpose our entire structure on a square grid of resolution 1/40.
This means that every circle will contain at least one grid point.   For each circle we choose
such a point as a Steiner vertex.   Next we join the Steiner vertices in the order of the circles
appearing on the track. This creates a polygon $P' = \{a_1, a_2, \ldots, a_{(O_n)}\}$.

The size of the circles is small enough that any polygon connecting them will not intersect $P_1$.  
An equivalent statement is to say that the convex hull of two consecutive circles avoids $P_1$.
This is trivially true for the rectangular components, and for any bend where the smaller arc is on $P_2$.
On the other hand if the bend's greater arc is on $P_2$, the only edge of the hull that could
intersect $P_1$ is the segment $s$ tangent to the circles, closest to $P_1$.
Consider the circular arc $c$ that is 1/4 of the way from $P_2$ to $P_1$. The two circles fit
tightly in the sub-tube between $c$ and $P_2$.  We place two segments, tangent to the two
circles in question, and orthogonal to $c$.  Let $x_1$ and $x_2$ be the intersection points
of the segments with $c$.  Every point on the segment between $x_1$ and $x_2$ 
is closer to $P_1$ than $s$ is.   Thus it suffices to establish that this segment does not
intersect $P_1$.   The arc corresponding to this segment is the angle spanned by three
circles.   As established earlier, this angle is a constant (roughly $\frac{3}{25}\pi$).
Thus the closest point of the segment to the lesser arc of the bend
is also a constant.
%
Thus the separation between  $P$ and $P'$ is at least 1/5.

Furthermore, it can be verified that no two Steiner vertices within a bend will have a separation
smaller than that between consecutive vertices.
What  remains is to prove that no two bends are close to each
other (in other words, we must prove that the Steiner vertices of $P'$ are sufficiently separated).
Every bend is associated with a vertex of $P$, or, to put it differently, with two consecutive edges of $P$.   We know that every Steiner vertex of a bend has distance at most 2/5 from both
associated edges of $P$.   When we compare Steiner vertices from two different bends,
we can identify one associated edge from each bend, such that the chosen edges are 
entirely disjoint (i.e. they don't even have a common endpoint).
Since those two edges must have a separation of at least 1, we can conclude that the 
two groups of Steiner vertices have separation at least 1/5.

At this point, we can conclude that the union of $P$ and $P'$ has a \rmfs\ of $\Theta(1)$.
We can also add chords from each $p_i$ to all the Steiner vertices in its associated bend.
This creates several triangles that are roughly isosceles and have a common apex.
Since the angle of each such triangle at the common apex is a constant, the separation
between the newly introduced chords and their opposite vertices is also constant.
As a result, the zone $Z$ between $P$ and $P'$ becomes quadrangulated, with
constant feature size.

Let $T$ be the trapezoidal decomposition of $P'$ resulting from extending horizontal rays from each of its vertices.   The construction of $T$ places an additional $O(n)$ Steiner vertices on $P'$
(i.e. on the inner boundary of $Z$).
The \rmfs\ of $T$ is $\Theta(1)$ since all its vertices are on our grid of resolution 1/40.
Horizontal lines are 1/40 apart, which means all new vertices are at least 1/40 apart from
other vertices, old or new.   
Since the new vertices are on $P'$, they are at a safe distance from the boundary of $P$ as well.
Thus we conclude the following.

\begin{theorem}
The planar straight line drawing consisting of the union of $Z$ and $T$ is a non-proper
quadrangulation with a constant  \rmfs.
\end{theorem}
\section*{Acknowledgments}

The research presented in this paper was initiated at the $24^{\mathrm{th}}$~Annual Winter Workshop on Computational Geometry at the Bellairs Research Institute of McGill University. The authors would like to thank  Godfried Toussaint as well as the non-author participants of the workshop for their helpful comments and for creating an environment conducive to creative thought:
Zachary Abel,
Brad Ballinger,  
 Nadia Benbernou, 
 Jit Bose, 
 Jean Cardinal, 
 Sebastien Collette, 
 Mirela Damian,  
 Marty Demaine, 
 Robin Flatland, 
 Ferran Hurtado,
 Scott Kominers, 
 Stefan Langerman, 
 Robbie Schweller,  
 David Wood, and
 Stefanie Wuhrer.

\bibliographystyle{plain}

\begin{thebibliography}{1}

\bibitem{Dey-2007}
Tamal~K. Dey.
\newblock Delaunay mesh generation of three dimensional domains.
\newblock Technical Report OSU-CISRC-09/07-TR64, Ohio State University, October
  2007.

\bibitem{Erickson-2003}
Jeff Erickson.
\newblock Nice point sets can have nasty delaunay triangulations.
\newblock {\em Discrete \& Computational Geometry}, 30(1), July 2003.

\bibitem{Frederickson-1997}
Greg~N. Frederickson.
\newblock {\em Dissections: Plane and Fancy}.
\newblock Cambridge University Press, November 1997.

\bibitem{Garey-Johnson-1979}
Michael~R. Garey and David~S. Johnson.
\newblock {\em Computers and Intractability: {A} Guide to the Theory of
  {NP}-Completeness}.
\newblock W. H. Freeman \& Co., 1979.

\bibitem{Hudson-Miller-Phillips-2006}
Benoit Hudson, Gary Miller, and Todd Phillips.
\newblock Sparse voronoi refinement.
\newblock In {\em Proceedings of the 15th International Meshing Roundtable},
  pages 339--356, Birmingham, Alabama, September 2006. Springer-Verlag.

\bibitem{Lowry-1814}
Mr. Lowry.
\newblock Solution to question 269, [proposed] by {Mr. W. Wallace}.
\newblock In T.~Leybourn, editor, {\em Mathematical Repository}, volume~3,
  part~1, pages 44--46. W. Glendinning, London, 1814.

\bibitem{Ruppert-1993}
Jim Ruppert.
\newblock A new and simple algorithm for quality 2-dimensional mesh generation.
\newblock In {\em Proceedings of the 4th Annual ACM-SIAM Symposium on Discrete
  Algorithms}, pages 83--92, Austin, Texas, 1993.

\end{thebibliography}

\end{document}